\documentclass[aps,prl,twocolumn]{revtex4}

\usepackage{graphicx}
\usepackage{dcolumn}
\usepackage{bm}

\begin{document}

\title{Second-quantized Landau-Zener theory for dynamical instabilities}
\author{J.R. Anglin}
\affiliation{Center for Ultracold Atoms, Massachusetts Institute of Technology, 77 Massachusetts Ave., Cambridge MA 02139}
\date{\today}

\begin{abstract}
State engineering in nonlinear quantum dynamics sometimes may demand driving
the system through a sequence of dynamically unstable intermediate states.
This very general scenario is especially relevant to dilute Bose-Einstein
condensates, for which ambitious control schemes have been based on the
powerful Gross-Pitaevskii mean field theory. Since this theory breaks down
on logarithmically short time scales in the presence of dynamical
instabilities, an interval of instabilities introduces quantum corrections,
which may possibly derail a control scheme. To provide a widely applicable
theory for such quantum corrections, this paper solves a general problem of
time-dependent quantum mechanical dynamical instability, by modelling it as
a second-quantized analogue of a Landau-Zener avoided crossing: a `twisted
crossing'.
\end{abstract}

\maketitle

Quantum dynamical instabilities (complex excitation frequencies) have been
associated with formation of solitons and vortices in driven Bose-Einstein
condensates. For example, in Refs. \cite{bh} and \cite{sc}, the Bogoliubov
spectra are examined for two-parameter families of mean field states.
Although the physics involved is significantly different (one-dimensional
modulated current versus rotating harmonic trap), figures in both these
works show dynamical instabilities in narrow, finger-like regions of
parameter space (see Figure 1). We will show below that this is generic for
dynamically unstable quasiparticle excitations in weakly interacting
many-body systems.

Such systems, especially condensates, may be sufficiently weakly damped that
state engineering under energetic instability becomes feasible. Interesting
structures such as solitons and sonic event horizons will require this
capability. But to access the entire volume of energetically unstable
control space, it will also be a common task to drive systems briefly
through narrow regions of dynamical instability. Mean field theory and
adiabatic approximations, which are important tools of state engineering,
will break down during these episodes \cite{VA}. This paper therefore
exploits an analogy between dynamical instability in second quantization,
and avoided crossing in two-state quantum mechanics \cite{LZ}, to determine
the quantum effects of driving a system through dynamical instability. An
almost identical calculation has recently been presented \cite{them} to deal
with the specific dynamical instability involved in dissociation of a
molecular condensate \cite{us}; the present paper provides an explicit
derivation of the result in a general context. 
\begin{figure}[tbp]
\includegraphics[width=.475\textwidth]{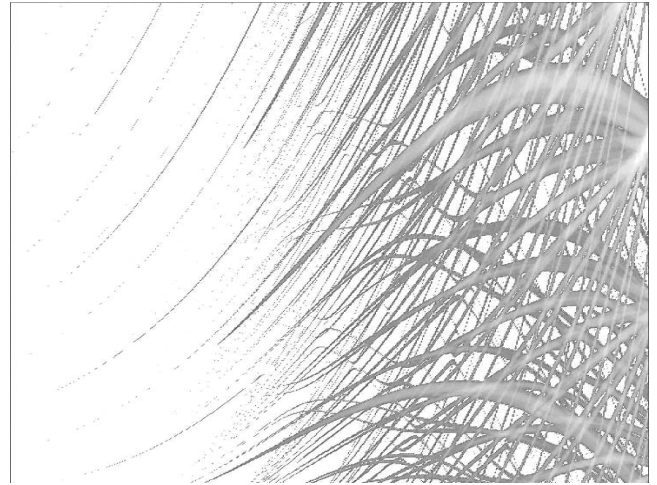}
\caption{Grayscale plot of sum of absolute values of imaginary parts of
Bogoliubov excitation frequencies for a quasi-one-dimensional condensate
flowing with winding number 10 around a toroidal trap with modulated
potential. Axes are control parameters: mean condensate density increases
from bottom to top, severity of density modulation increases from left to
right. Compare to figures in Refs. \protect\cite{bh,sc}.}
\end{figure}

Quantum dynamical instability is not always recognized as a possibility,
because with a Hermitian Hamiltonian, the eigenfrequencies of
Schr\"{o}dinger evolution must all be real. But the eigenfrequencies of
Heisenberg evolution may be complex. It is easy to see this for the
manifestly Hermitian Hamiltonian 
\begin{eqnarray}
\hat{H} &=&\left( \omega +\delta \right) \hat{a}_{+}^{\dagger }\hat{a}%
_{+}+\left( \delta -\omega \right) \hat{a}_{-}^{\dagger }\hat{a}_{-} 
\nonumber \\
&&+\gamma \left( \hat{a}_{+}^{\dagger }\hat{a}_{-}^{\dagger }+\hat{a}_{+}%
\hat{a}_{-}\right)  \label{H1}
\end{eqnarray}
where all co-efficients are real, and $\hat{a}_\pm$ and $\hat{a}_\pm^\dagger$
are ordinary annihilation and creation operators (\textit{i.e.} $[\hat{a}%
_{j},\hat{a}_{k}^{\dagger }] =\delta _{jk}$ is the only non-vanishing
commutator). The eigenoperators of Heisenberg evolution satisfy 
\begin{eqnarray}
i\dot{\hat{b}}_{\pm} &=&[\hat{b}_\pm,\hat{H}] =\Omega_\pm\hat{b}_\pm
\label{eigen} \\
\hat{b}_\pm &=&\frac{(\delta \pm \sqrt{\delta^2-\gamma^2})^{\frac{1}{2}}\hat{%
a}_+ +(\delta \mp \sqrt{\delta^2-\gamma^2})^{\frac{1}{2}}\hat{a}_-^\dagger} {%
\sqrt{2}|\delta^2-\gamma^2|^{1/4}}  \label{apm} \\
\Omega _\pm &=&\omega \pm \sqrt{\delta^2-\gamma^2}  \label{omegpm}
\end{eqnarray}
where we assume $\gamma \geq 0$ (without loss of generality, since we can
always redefine $\hat{a}_+\to-\hat{a}_+$). These operators always obey $[%
\hat{b}_{+},\hat{b}_{-}]=0$. As long as $\gamma <\left| \delta \right| $, $%
\Omega _{\pm }$ are both real, and we also have 
\begin{equation}
\gamma < |\delta| :\quad [\hat{b}_\pm,\hat{b}_\pm^\dagger] =\pm\mathrm{sgn}%
(\delta)\quad , \quad [\hat{b}_\pm,\hat{b}_\mp^\dagger] =0\ .
\end{equation}
And we can write 
\begin{equation}
\hat{H}=\mathrm{sgn}(\delta)\times\left[E_0 +\sum_\pm \pm\Omega_\pm \hat{b}%
_\pm^\dagger\hat{b}_\pm\right]\ ,\qquad \gamma<|\delta|  \label{Omegareal}
\end{equation}
for $E_0=\delta -\omega $. This means that as long as $\gamma < |\delta|$,
one of the $\hat{b}_\pm$ is really a creation operator which would have a $%
^\dagger$ in more conventional notation; but apart from this trivial issue
of labelling, we have diagonalized to noninteracting quasiparticles of the
standard (bosonic) kind.

If $\gamma >|\delta|$, however, $\Omega _\pm$ are complex, and neither $\hat{%
b}_\pm$ nor $\hat{b}_\pm^\dagger$ are standard annihilation operators. They
obey 
\begin{equation}
\gamma > |\delta| :\quad [\hat{b}_\pm,\hat{b}_\pm^\dagger] =0\quad , \quad [%
\hat{b}_\pm,\hat{b}_\mp^\dagger] =\pm i
\end{equation}
so that we must say that the \textit{canonical} conjugate of $\hat{b}_\pm$
is $\mp i\hat{b}_\mp^\dagger$, and no longer coincides with the Hermitian
conjugate. In this case we have 
\begin{equation}
\hat{H}=E_0+\sum_\pm \mp i\Omega_\pm \hat{b}_\mp^\dagger \hat{b}_\pm\
,\qquad \gamma>|\delta|  \label{Omegacomplex}
\end{equation}
instead of (\ref{Omegareal}). One can also show that if $\gamma >|\delta |$
then the eigenvalues of $\hat{H}$ no longer involve integer multiples of $%
\Omega _{\pm }$: they remain real, but are continuous. Hence in this case
there is no way to interpret the energy eigenstates as having definite
numbers of any kind of quasiparticle.

If $\gamma >|\delta |$, then $\hat{H}$ is unbounded below as well as above;
this can also occur even for $\gamma <|\delta |$, if $|\omega| > \sqrt{%
\delta^2-\gamma^2}$. In these cases, (\ref{H1}) can be physical only as the
linearization of a nonlinear Hamiltonian about an excited state. And in all
cases, since the `anomalous coupling' term proportional to $\gamma $ in (\ref
{H1}) does not conserve the numbers of particles created and annihilated by $%
\hat{a}_\pm^\dagger$ and $\hat{a}_\pm$, these must be some kind of
quasiparticle, or else there must be a source of atoms (such as a molecular
condensate treated as a c-number field \cite{them}). But there are many
realistic situations in which unboundedness of the linearized Hamiltonian,
and non-conservation of (quasi-)particles, are perfectly correct. For these (%
\ref{H1}) is universal, in the sense that for all eigenfrequencies, whether
real or complex, pairs of modes may be expressed in this form by using (\ref
{apm}) to define $\hat{a}_{\pm}$.

Two-body interactions among particles lead to quasiparticle Hamiltonians
with anomalous couplings as in (\ref{H1}). Since $\gamma $ is thus
constrained to be small if interactions are weak, dynamical instabilities
tend to arise in weakly interacting systems not through $\gamma $ becoming
large, but through $\delta $ becoming even smaller. A dynamical instability
is thus a kind of second-quantized analogue to an avoided crossing \cite{him}%
. See Figure 2. If the many-body Hamiltonian depends on some number $D$ of
control parameters, then any two modes will have frequencies exactly
opposite on a $(D-1)$-dimensional parameter subspace, and so there will be a
dynamical instability within a thin shell around that (hyper)surface. Of
course, dynamical instabilities can only arise if there is energetic
instability, so that some negative frequency modes exist. But if the system
is energetically unstable, then finger-like (or in higher dimensioned
control spaces, hypershell-like) regions of dynamical instability are
generic. 
\begin{figure}[tbp]
\includegraphics[width=.475\textwidth]{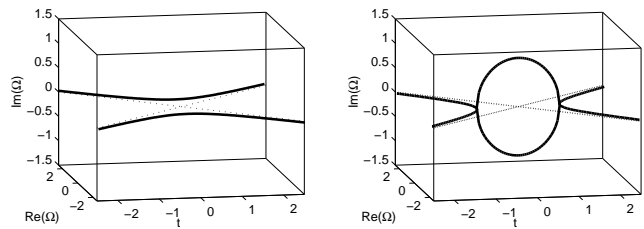}
\caption{At left, an ordinary avoided crossing; at right, a `twisted
crossing' with complex intermediate eigenfrequencies.}
\end{figure}

Since much state engineering can be accomplished by means of time-dependent
control parameters in the Hamiltonian, we need to assess the effects of a
time dependence that brings a system through an intermediate interval in
which one dynamical instability exists. Focusing our attention on the pair
of modes involved in the instability, we can always cast them in the form of
(\ref{H1}), with time-dependent $\omega ,\delta ,\gamma $. As long as $%
|\delta |\gg \gamma $ we can apply standard methods, but a connection
formula is required to give the quantum evolution over the intermediate
interval. Within this interval, we assume the widely applicable case that $%
\omega$ and $\gamma$ are essentially constant, and $\delta$ changes linearly
with time. Since $(\hat{a}_+^\dagger \hat{a}_+-\hat{a}_-^\dagger\hat{a}_-)$
commutes with $\hat{H}$, we can eliminate $\omega$ in an interaction picture
by taking $\hat{a}_\pm\to e^{\mp i\omega t}\hat{a}_\pm$, giving us the
problem 
\begin{equation}  \label{problem}
\hat{H}_{i}=\gamma (\hat{a}_+^\dagger\hat{a}_-^\dagger+\hat{a}_+\hat{a}_-)
+\zeta^2 t (\hat{a}_+^\dagger\hat{a}_+ + \hat{a}_-^\dagger\hat{a}_-)
\end{equation}
for a constant rate $\zeta$.

By examining the Heisenberg equations of motion with this time-dependent
Hamiltonian we can straightforwardly obtain 
\begin{eqnarray}
\lim_{t\rightarrow -\infty }\hat{a}_\pm(t) &=& e^{-i\theta(t)}\hat{a}_{\pm{\
IN}} -\frac{\gamma e^{i\theta (t)}}{2\zeta^{2}t}\hat{a}_{\mp {\ IN}}^\dagger
\label{IN} \\
\lim_{t\rightarrow +\infty }\hat{a}_\pm(t) &=&e^{-i\theta (t)}\hat{a}_{\pm {%
\ OUT}} -\frac{\gamma e^{i\theta (t)}}{2\zeta^{2}t}\hat{a}_{\mp {\ OUT}%
}^{\dagger}  \label{OUT}
\end{eqnarray}
where 
\begin{equation}
\theta(t)=\frac{\zeta^2 t^2}{2}-\frac{\gamma^2}{2\zeta^{2}}\ln|2\zeta t|+%
\mathcal{O}(t^{-2})  \label{theta}
\end{equation}
and $\hat{a}_{\pm IN,OUT}$ are time-independent annihilation operators,
whose relationship to each other unfortunately depends on the intervening
evolution at finite times. (The factor multiplying $t$ in the argument of
the logarithm in (\ref{theta}) is not actually fixed by the equation of
motion, but is chosen for our notational convenience below.) Of course the
point of the limits $t\rightarrow \pm \infty$ is not literally to consider
infinite times, but only to give the asymptotic behaviour which can be
smoothly matched to adiabatic approximations valid for $\zeta^2 |t| \gg
\gamma$. Hence in the Heisenberg picture no further definition of $\hat{a}%
_{\pm IN}$ is required: it will be given in terms of initial time operators
by solving the evolution problem for $t \ll -\zeta^{-2}\gamma$. So the
problem is to determine the $_{OUT}$ operators in terms of the $_{IN}$
operators.

This problem is directly analogous to the Landau-Zener problem for
amplitudes in a two-state system being driven through an avoided crossing,
except that the amplitudes are replaced by operators, and the
second-quantized problem allows the `anomalous' coupling which would be
non-Hermitian in the two-state system. Because our problem is also linear,
we can solve it in a very similar way. Iterating the Heisenberg equations to
obtain a second order equation for $\hat{a}_+$ with $\hat{a}_-^\dagger$
eliminated, 
\begin{equation}
\left[\partial_{tt}+2i\zeta^2t\partial_t-\gamma^2+2i\zeta^2\right]
\left(e^{-i\frac{\zeta^2 t^2}{2}}\hat{a}_+\right) =0
\end{equation}
then Fourier transforming from $t$ to frequency $\nu$, we find an
easily-solved first order differential equation in $\nu$: 
\begin{eqnarray}
\hat{a}_+(t) &\equiv& e^{i\frac{\zeta^2 t^2}{2}}\int d\nu \,e^{-i\nu t}\hat{%
\alpha}(\nu)  \nonumber \\
\partial_{\nu}\hat{\alpha} &=&-\frac{(\nu^2+\gamma^2)}{2i\zeta^2\nu}\hat{%
\alpha}\ .
\end{eqnarray}

Inverse Fourier transforming back to $t$ then gives 
\begin{eqnarray}
\hat{a}_{+}(t) &=&\sum_{j=1,2}\hat{\alpha}_{j}A_{j}(t)  \nonumber  \label{at}
\\
A_{j}(t) &\equiv &e^{i\frac{\zeta ^{2}t^{2}}{2}}\int\limits_{C_{j}}\frac{%
d\nu }{\zeta }\,e^{iS_{t}(\nu )}  \nonumber \\
S_{t}(\nu ) &\equiv &-\nu t+\frac{\nu ^{2}}{4\zeta ^{2}}+\frac{\gamma ^{2}}{%
2\zeta ^{2}}\ln \frac{\nu }{\zeta }\ .
\end{eqnarray}
There are two independent solutions $A_{1,2}(t)$, with independent
operator-valued co-efficients $\hat{\alpha}_{1,2}$, since the $\nu $
integral may be evaluated along inequivalent contours $C_{1,2}$ in the
complex $\nu $-plane. Since the integrand vanishes at the branch point $\nu
=0$, and at infinity between angles $(0,\frac{\pi }{2})$ and $(\pi ,\frac{%
3\pi }{2})$, integration contours can only terminate at these points
(because terminating anywhere else amounts to replacing the integrand with
something that does vanish elsewhere). The simplest distinct contours are $%
C_{1}$ running from lower-left infinity to upper-right, and $C_{2}$ running
from zero to upper-right infinity. Since the integrand is analytic except at
upper-left and lower-right infinity, and on the branch cut (which we take to
run along the negative imaginary axis), we are free to choose contours that
make the evaluation easier; see Fig.~3. 
\begin{figure}[tbp]
\includegraphics[width=.475\textwidth]{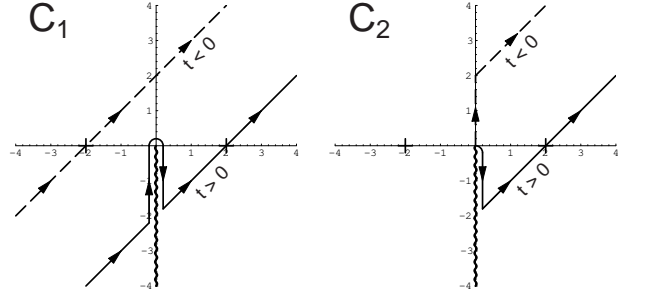}
\caption{Contours used in the text, for $\protect\zeta ^{2}t\gg \protect%
\gamma $ (solid) and $\protect\zeta ^{2}t\ll -\protect\gamma $ (dashed),
plotted in the complex plane of $\protect\nu /(\protect\zeta ^{2}|t|)$.
Saddlepoints are marked with crosses on the real axis at $2\,\mathrm{sgn}(t)$%
. At left, contour $C_{1}$ traverses from lower-left to upper-right
infinity; at right, contour $C_{2}$ runs from 0 to upper-right infinity.}
\end{figure}

A given choice of contour yields a single solution valid for all $t$, but
for $|t| \rightarrow \infty $ the integral may be evaluated using
combinations of the method of steepest descents and integral representations
of the Gamma function. (There is a second saddlepoint near the origin, but
as $|t| \rightarrow\infty$ it is too close to the branch point for the
method of steepest descents to work, and so contour $C_{2}$ must run along
the imaginary axis until it reaches a contour of steepest descent through
the other saddlepoint.) The asymptotic behaviours of the full solutions $%
A_{j}(t)$\ can thus be obtained explicitly, allowing us to determine the
late time behaviour of the solutions whose early time behaviour matches (\ref
{IN}).

The contributions to the contour integrals for $A_{j}(t)$\ from the diagonal
portions are dominated by the saddlepoints $\nu_{t}=2\zeta^{2}t +\mathcal{O}(%
\frac{\gamma^2}{\zeta^2 t})$, since for $\nu =2\zeta^2 t+e^{i\pi/4}\zeta s$, 
\begin{eqnarray}
\int\!\frac{d\nu}{\zeta}e^{iS} &=&\int\! ds\,e^{-{\frac{1}{4}}s^{2}} \left[
1+\frac{i\gamma^2e^{i\frac{\pi}{4}}s}{2\zeta^3 t} +\mathcal{O}(\frac{s^2}{%
\zeta^2 t^2}) \right]  \nonumber \\
&&\times e^{-i( \zeta^2 t^2-\frac{\gamma^2}{2\zeta^2}\ln(2\zeta t) -\frac{\pi%
}{4})}\ .
\end{eqnarray}
So if the contour passes through the saddlepoint we get a contribution to $%
A_j(t)$ 
\begin{eqnarray}
[1+\mathcal{O}(\zeta t)^{-2}] A_{SP} &=& 2\sqrt{\pi}e^{i\frac{\pi}{4}}
e^{-i\left({\frac{\zeta^2t^2}{2}}-{\frac{\gamma^2}{2\zeta^2}}\ln|2\zeta
t|\right)}  \nonumber \\
&&\times \left\{
\begin{array}{ccc}
1 & , & t>0 \\ 
e^{-\frac{\pi\gamma^2}{2\zeta^2}} & , & t<0
\end{array}
\right.\ .
\end{eqnarray}
If a diagonal segment shown in Fig.~3 does not cross a saddlepoint, then
this segment yields only an utterly negligible contribution $\sim
e^{-2\zeta^2 t^2}$.

For the vertical segments along the imaginary axis in $C_2$, we let $\nu =
-i\xi/t+0$ and integrate 
\begin{eqnarray}
\int\!{\frac{d\nu}{\zeta}}e^{iS_t}&=&\frac{1}{i\zeta t}\int_0^{2(\zeta
t)^2}d\xi\, e^{-\xi+i{\frac{\gamma^2}{2\zeta^2}}\ln{\frac{0-i\xi}{\zeta t}}}
e^{-i{\frac{\xi^2}{4\zeta^2t^2}}}  \nonumber \\
&\doteq&\frac{e^{-i{\frac{\gamma^2}{2\zeta^2}}\ln|\zeta t|} e^{{\frac{%
\pi\gamma^2}{4\zeta^2}}\mathrm{sgn}(t)}}{i\zeta t} \int_0^\infty\!d\xi
\,\xi^{\frac{i\gamma^2}{2\zeta^2}}e^{-\xi}  \nonumber \\
&=&\frac{e^{-i{\frac{\gamma^2}{2\zeta^2}}\ln|\zeta t|} e^{{\frac{\pi\gamma^2%
}{4\zeta^2}}\mathrm{sgn}(t)}}{i\zeta t} \Gamma\left(1+\frac{i\gamma^2}{%
2\zeta^2}\right)
\end{eqnarray}
where in the second line we extend the integration limit to infinity by
ignoring terms of order $e^{-2(\zeta t)^2}$, and drop the $\exp-i{\frac{\xi^2%
}{4\zeta^2t^2}}$ from the integrand because it yields corrections of order $%
(\zeta t)^{-2}$. This yields the contribution 
\begin{eqnarray}  \label{Aim}
A_{IM}&=&{\frac{2^{i{\frac{\gamma^2}{2\zeta^2}}}}{i\zeta t}} e^{{\frac{%
\pi\gamma^2}{4\zeta^2}}\mathrm{sgn}(t)} \Gamma\left(1+\frac{i\gamma^2}{%
2\zeta^2}\right) e^{i[{\frac{\zeta^2t^2}{2}}-{\frac{\gamma^2}{2\zeta^2}}%
\ln|2\zeta t|]}  \nonumber \\
&&\times[1+\mathcal{O}(\zeta t)^{-2}] \ .
\end{eqnarray}
From the two segments at $t>0$ where $C_1$ hugs the branch cut, we gain a
contribution $(1-e^{-{\frac{\pi\gamma^2}{\zeta^2}}})A_{IM}$.

Combining these contributions and recalling our definition of $\theta(t)$ in
(\ref{theta}), we obtain, up to corrections of order $(\zeta t)^{-2}$, 
\begin{eqnarray}  \label{A12}
A_{1}(t) &=&\left\{
\begin{array}{ccc}
2\sqrt{\pi}e^{i\frac{\pi}{4}}e^{-i\theta(t)}e^{-\frac{\pi\gamma^2}{2\zeta^2}}
& , & t<0 \\ 
2\sqrt{\pi}e^{i\frac{\pi}{4}}e^{-i\theta(t)}+\mathcal{O}(\zeta t)^{-1} & , & 
t>0
\end{array}
\right. \\
A_{2}(t) &=&\left\{ 
\begin{array}{ccc}
{\frac{2^{i{\frac{\gamma^2}{2\zeta^2}}}}{i\zeta t}}e^{-{\frac{\pi\gamma^2}{%
4\zeta^2}}}e^{i\theta(t)} \Gamma\left(1+\frac{i\gamma^2}{2\zeta^2}\right) & ,
& t<0 \\ 
2\sqrt{\pi}e^{i\frac{\pi}{4}}e^{-i\theta(t)} +\mathcal{O}(\zeta t)^{-1} & ,
& t>0
\end{array}
\right.\ ,
\end{eqnarray}
where the suppressed $\mathcal{O}(\zeta t)^{-1}$ terms can easily be
computed from (\ref{Aim}), but will not be needed.

Comparing (\ref{A12}) with (\ref{IN}) and (\ref{at}), we can read off 
\begin{eqnarray}
\hat{\alpha}_{1} &=& {\frac{e^{-i{\frac{\pi}{4}}} e^{\frac{\pi\gamma^2}{%
2\zeta^2}}}{2\sqrt{\pi}}}\hat{a}_{+IN} \\
\hat{\alpha}_{2} &=& {\frac{2^{-i{\frac{\gamma^2}{2\zeta^2}}}i\gamma}{2\zeta}%
} {\frac{e^{\frac{\pi\gamma^2}{4\zeta^2}}}{\Gamma\left(1+\frac{i\gamma^2}{%
2\zeta^2}\right)}} \hat{a}^\dagger_{-IN}  \nonumber \\
&=&{\frac{2^{-i{\frac{\gamma^2}{2\zeta^2}}}}{2i\sqrt{\pi}}} \left[{\frac{%
\Gamma(1-{\scriptstyle{\frac{i\gamma^2}{2\zeta^2}}})}{\Gamma(1+{\scriptstyle{%
\frac{i\gamma^2}{2\zeta^2}}})}}\right]^{\frac{1}{2}} \left[e^{\frac{%
\pi\gamma^2}{\zeta^2}}-1\right]^{\frac{1}{2}}\hat{a}^\dagger_{-IN}
\end{eqnarray}
using the $\Gamma$-function identity 
\begin{equation}
|\Gamma(1+ix)|^2={\frac{\pi x}{\sinh\pi x}}\ .
\end{equation}
Then comparing these last results with (\ref{OUT}), and noting that the
equations for $\hat{a}_-$ and $\hat{a}^\dagger_+$ are exactly similar to
those we have examined, we finally conclude 
\begin{eqnarray}
\hat{a}_{\pm {\ OUT}} &=&e^{\frac{\pi}{2}\gamma^2/\zeta^2}\hat{a}_{\pm{\ IN}%
} +e^{i\eta}\left[e^{\pi\gamma^2/\zeta^2}-1\right]^{\frac{1}{2}}\hat{a}_{\mp 
{\ IN}}^\dagger  \nonumber \\
e^{i\eta} &=&2^{-i{\frac{\gamma^2}{2\zeta^2}}}e^{-i{\frac{\pi}{4}}} \left[{%
\frac{\Gamma\left(1-{\frac{i\gamma^2}{2\zeta^2}}\right)}{\Gamma\left(1+{%
\frac{i\gamma^2}{2\zeta^2}}\right)}}\right]^{\frac{1}{2}}\ .  \label{main}
\end{eqnarray}

Equation (\ref{main}) is our main result. It is indeed very similar to the
Landau-Zener formula, except for the vital difference that the exponents are
all positive rather than negative. Consequently, in the state which is
annihilated by $\hat{a}_{\pm IN}$, the expected numbers of quasiparticles
after passing through the instability will be 
\begin{equation}
\left\langle 0_{{\ IN}}\left| \hat{a}_{\pm{\ OUT}}^\dagger\hat{a}_{\pm {\ OUT%
}} \right| 0_{{\ IN}}\right\rangle = e^{\frac{\pi\gamma^2}{\zeta^2}}-1\ .
\end{equation}
Driving through a dynamical instability in less than the characteristic time
scale of the instability produces only mild excitation, but exceeding this
time generates quasiparticles explosively. And the crossover between these
regimes is rather abrupt. This may set important practical limits to the
precision of state engineering, because controlling faster than $\gamma$ may
mean not being slow enough to be adiabatic for other, dynamically stable
modes of the total nonlinear system.

Note that changing $\zeta^2\to -\zeta^2$ in (\ref{problem}), so that the
system is stable at early times rather than late, produces only $%
e^{i\eta}\to -e^{-i\eta}$ in (\ref{main}). And while we have analyzed the
most general type of dynamical instability, with complex frequency and two
degrees of freedom involved, the simpler case with purely imaginary
frequency, which was treated in \cite{them}, can be obtained by setting $%
\omega\to0$, and $\hat{a}_\pm\to\hat{a}/\sqrt{2}$ in (\ref{H1}). Everywhere
else, one must use simply $\hat{a}_\pm\to\hat{a}$ instead; there are also
trivial changes to (\ref{Omegareal}) and (\ref{Omegacomplex}). We have
therefore obtained a quite general result that will be useful for state
engineering through dynamical instabilities.


\begin{thebibliography}{9}
\bibitem{bh}  L.J. Garay, J.R. Anglin, J.I. Cirac and P. Zoller, Phys. Rev.
Lett.~\textbf{85}, 4643 (2000); Phys. Rev. \textbf{A63}, 023611 (2001).

\bibitem{sc}  S. Sinha and Y. Castin, Phys. Rev. Lett. \textbf{87}, 190402
(2001).

\bibitem{VA}  A. Vardi and J.R. Anglin, Phys. Rev. Lett. \textbf{86}, 568
(2001); J.R. Anglin and A. Vardi, Phys. Rev. \textbf{A64}, 013605 (2001).

\bibitem{LZ}  L.D. Landau, Phys. Z. Sowjetunion \textbf{2}, 46 (1932); C.
Zener, Proc. R. Soc. London A \textbf{137}, 696 (1932).

\bibitem{them}  V.A. Yurovsky, A. Ben-Reuven, and P.S. Julienne,
cond-mat/0108372.

\bibitem{us}  A. Vardi, V.A. Yurovsky and J.R. Anglin, Phys. Rev. \textbf{A64%
}, 063611 (2001).

\bibitem{him}  D.V. Skryabin, Phys. Rev. \textbf{A63}, 013602 (2001).
\end{thebibliography}
\end{document}